\documentclass[final]{svjour2}
\usepackage{graphicx}
\usepackage{rotating}
\usepackage{amssymb}
\usepackage{mathptmx}
\usepackage[numbers]{natbib}
\makeatletter
\journalname{Journal of Low Temperature Physics}
\bibpunct{}{}{,}{s}{}{,}

\begin{document}

\newcommand{\hdblarrow}{H\makebox[0.9ex][l]{$\downdownarrows$}-}
\title{The Electron Capture $^{163}$Ho Experiment ECHo}

\author{L. Gastaldo \and K. Blaum \and A. Doerr \and Ch. E. D$\ddot{u}$llmann \and K. Eberhardt \and S. Eliseev \and C. Enss \and Amand Faessler \and A. Fleischmann \and S. Kempf \and M. Krivoruchenko \and S. Lahiri \and M. Maiti \and Yu. N. Novikov \and P. C.-O. Ranitzsch \and F. Simkovic \and Z. Szusc \and M. Wegner}

\institute{L. Gastaldo, C. Enss, A. Fleischmann, S. Kempf, P. C.-O. Ranitzsch, M. Wegner \at
Kirchhoff Institute for Physics, Heidelberg University, INF 227 D-69120 Heidelberg, Germany,
Tel.: +49 6221 549886 \\
Fax: +49 6221 5498\\
\email{Loredana.Gastaldo@kip.uni-heidelberg.de}\\
http://www.kip.uni-heidelberg.de/echo/
\and
K. Blaum, A. Doerr, S. Eliseev \at
Max-Planck Institute for Nuclear Physics, Saupfercheckweg 1, 69117, Heidelberg Germany
\and
Ch. E. D$\ddot{u}$llmann \at
Institute for Nuclear Chemistry, Johannes Gutenberg University, Fritz Strassmann Weg 2, D-55128, Mainz, Germany,\\
 GSI Helmholtzzentrum f$\ddot{u}$r Schwerionenforschung, Planckstra${\ss}$e 1, 64291, Darmstadt, Germany,\\
 Helmholtz Institute Mainz, Johann-Joachim-Becherweg 36, 55128 Mainz, Germany
\and
K. Eberhardt \at
Institute for Nuclear Chemistry, Johannes Gutenberg University, Fritz Strassmann Weg 2, D-55128, Mainz, Germany,\\
 Helmholtz Institute Mainz, Johann-Joachim-Becherweg 36, 55128 Mainz, Germany
\and
Amand Faessler \at
Institute for Theoretical Physics, University of T$\ddot{u}$bingen, Morgenstelle 14; D-72076 T$\ddot{u}$bingen, Germany
\and
M. Krivoruchenko \at
Institute for Theoretical and Experimental Physics, B. Cheremushkinskaya 25, 117259 Moscow, Russia
\and
S. Lahiri \at
Saha Institute of Nuclear Physics, 1/AF Bidhannagar, 700064 Kolkata, India
\and
M. Maiti \at
Department of Physics, Indian Institute of Technology Roorkee, Roorkee-247667, India
\and
Yu. N. Novikov \at
Petersburg Nuclear Physics Institute, 188300 Gatchina, Russia
\and
F. Simkovic \at
Department of Nuclear Physics and Biophysics, Comenius University, Mlynska dolina, pav. F1; SK-84248 Bratislava, Slovakia
\and
Z. Szcus \at
Department of Cyclotron Laboratory,
Institute of Nuclear Research of the Hungarian Academy of Sciences,
4026 Debrecen, Bem ter 18/C, HUNGARY
}

\date{XX.XX.20XX}

\maketitle

\begin{abstract}

The determination of the absolute scale of the neutrino masses is one of the most challenging present questions in particle physics. The most stringent limit, $m(\bar{\nu}_{\mathrm{e}})<2$eV, was achieved for the electron anti-neutrino mass \cite{numass}. Different approaches are followed to achieve a sensitivity on neutrino masses in the sub-eV range. Among them, experiments exploring the beta decay or electron capture of suitable nuclides can provide information on the electron neutrino mass value. We present the Electron Capture $^{163}$Ho experiment ECHo, which aims to investigate the electron neutrino mass in the sub-eV range by means of the analysis of the calorimetrically measured energy spectrum following electron capture of $^{163}$Ho. A high precision and high statistics spectrum will be measured with arrays of metallic magnetic calorimeters. We discuss some of the essential aspects of ECHo to reach the proposed sensitivity: detector optimization and performance, multiplexed readout, $^{163}$Ho source production and purification, as well as a precise theoretical and experimental parameterization of the calorimetric EC spectrum including in particular the value of $Q_{\mathrm{EC}}$. We present preliminary results obtained with a first prototype of single channel detectors as well as a first 64-pixel chip with integrated micro-wave SQUID multiplexer, which will already allow to investigate $m(\nu_{\mathrm{e}})$ in the eV range.

\keywords{Neutrino mass, Metallic magnetic calorimeters, $^{163}$Ho}

\end{abstract}

\section{Introduction}

\noindent The fact that neutrinos are massive particles has been accepted since the discovery of neutrino flavor oscillations. Experiments investigating neutrino oscillations with solar neutrinos, reactor neutrinos, atmospheric neutrinos and neutrinos produced at accelerator facilities have been able to precisely define the three mixing angles $\theta_{12}$, $\theta_{23}$ and $\theta_{13}$, and the difference in the squared mass eigenvalues $\Delta m^2_{21}$ and $\Delta m^2_{31}$ \cite{Fogli_2012}.
While future improved neutrino oscillation experiments will provide information about the hierarchy of the mass eigenvalues, they will not be able to set the mass energy scale. The determination of the neutrino mass energy scale is a complicated task due to the weak interaction of neutrinos and to the smallness of their mass. There exist several approaches which can potentially yield this result: the analysis of the visible structures in the universe \cite{cosmology}, the existence of neutrinoless double beta decay \cite{0v2b}, the measurement of the time of flight of neutrinos emitted in Supernova explosions \cite{supernova} and the analysis of the kinematics of low energy beta decays and electron capture (EC) processes \cite{Drexlin_2013}. Presently the most stringent limit, $m(\bar{\nu}_{\mathrm{e}})<2$eV, was achieved for the electron anti-neutrino mass \cite{numass} by the analysis of the endpoint region of the $^3$H beta spectrum.

\noindent The Electron Capture $^{163}$Ho Experiment, ECHo, has the aim to investigate the electron neutrino mass in the energy range below 1 eV by a high precision and high statistics calorimetric measurement of the $^{163}$Ho electron capture spectrum \cite{De_Rujula_1982}, \cite{De_Rujula_2013}. Among all the nuclides undergoing electron capture processes, $^{163}$Ho has the lowest energy available to the decay $Q_{\mathrm{EC}}\,\sim \, 2.5\,$keV, which is given by the difference in the mass of mother and daughter. This is the reason why presently $^{163}$Ho is the best candidate to perform experiments investigating the value of the electron neutrino mass. On the other hand the very low $Q_{\mathrm{EC}}$ implies that detectors showing high energy resolution in the energy range below $3\,$keV need to be used. Presently the detectors that can measure energies below 3 keV with the highest precision are low temperature micro-calorimeters \cite{Enss_book}. Within the ECHo experiment, low temperature metallic magnetic calorimeters (MMCs) will be used. %In the next session a brief description of MMCs and the results achieved with a first prototype of MMCs having gold absorbers with $^{163}$Ho ion implanted will be presented.
In order to reach the aimed sensitivity on the electron neutrino mass, a number of experimental techniques and investigations have to be brought together, not only the development high energy resolution and fast detectors: $^{163}$Ho source production and purification,  precise theoretical and experimental parameterization of the calorimetric EC spectrum including in particular the value of $Q_{\mathrm{EC}}$ which should be precisely measured independently. In the following a short discussion on the present investigations within the ECHo experiment will be reported.

\section{First prototype Metallic Magnetic Calorimeters with implanted $^{163}$Ho}

\begin{figure}[t]
\minipage{0.45\textwidth}
\begin{center}
\includegraphics[width=1\textwidth]{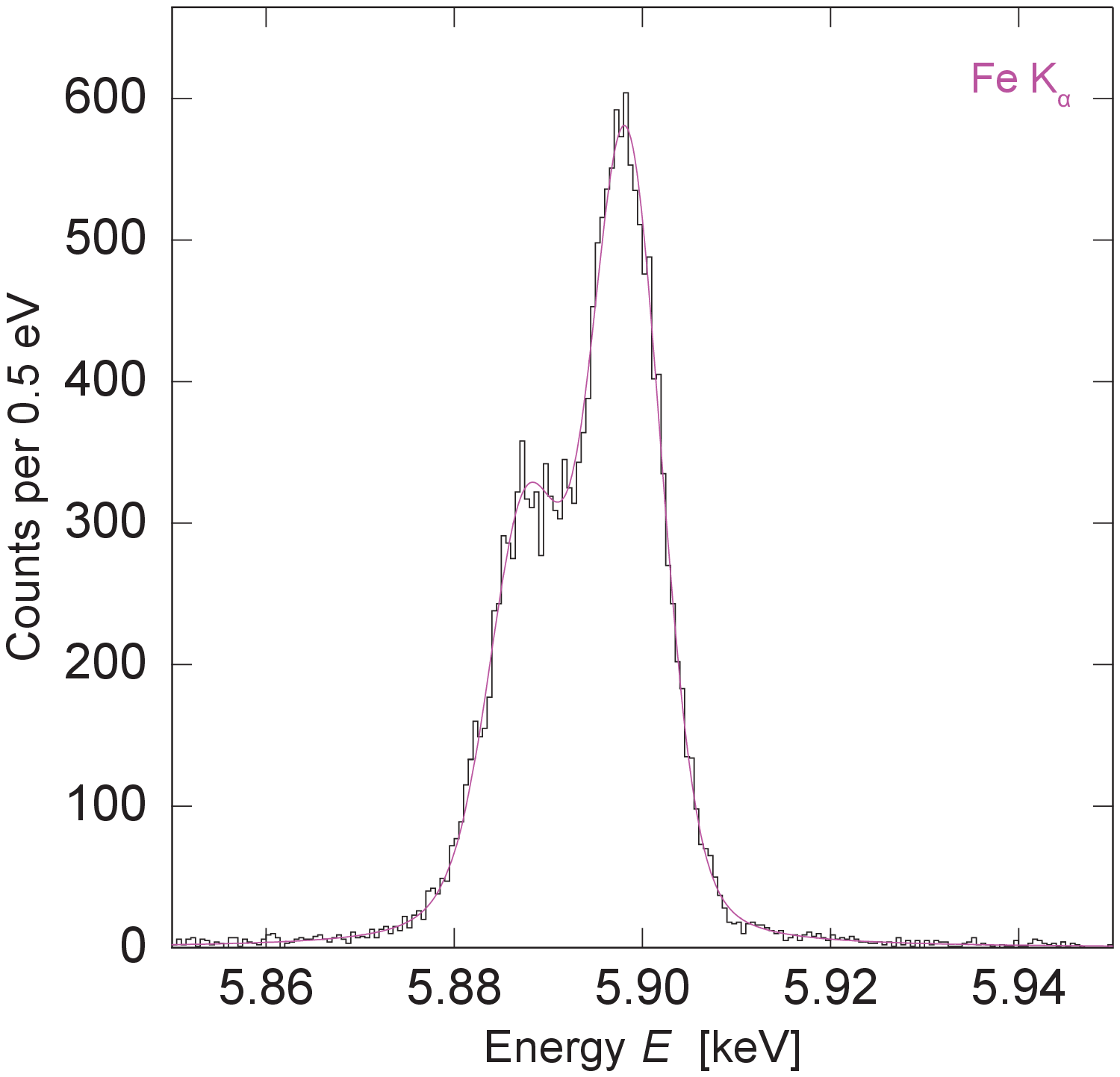}
%\caption{Methods}
%\label{fg:methods}
\end{center}
\endminipage\hfill
%\minipage{0.1\textwidth}
%\begin{center}
%\end{center}
%\endminipage\hfill
\minipage{0.5\textwidth}
\begin{center}
\includegraphics[width=0.85\textwidth, angle=-90]{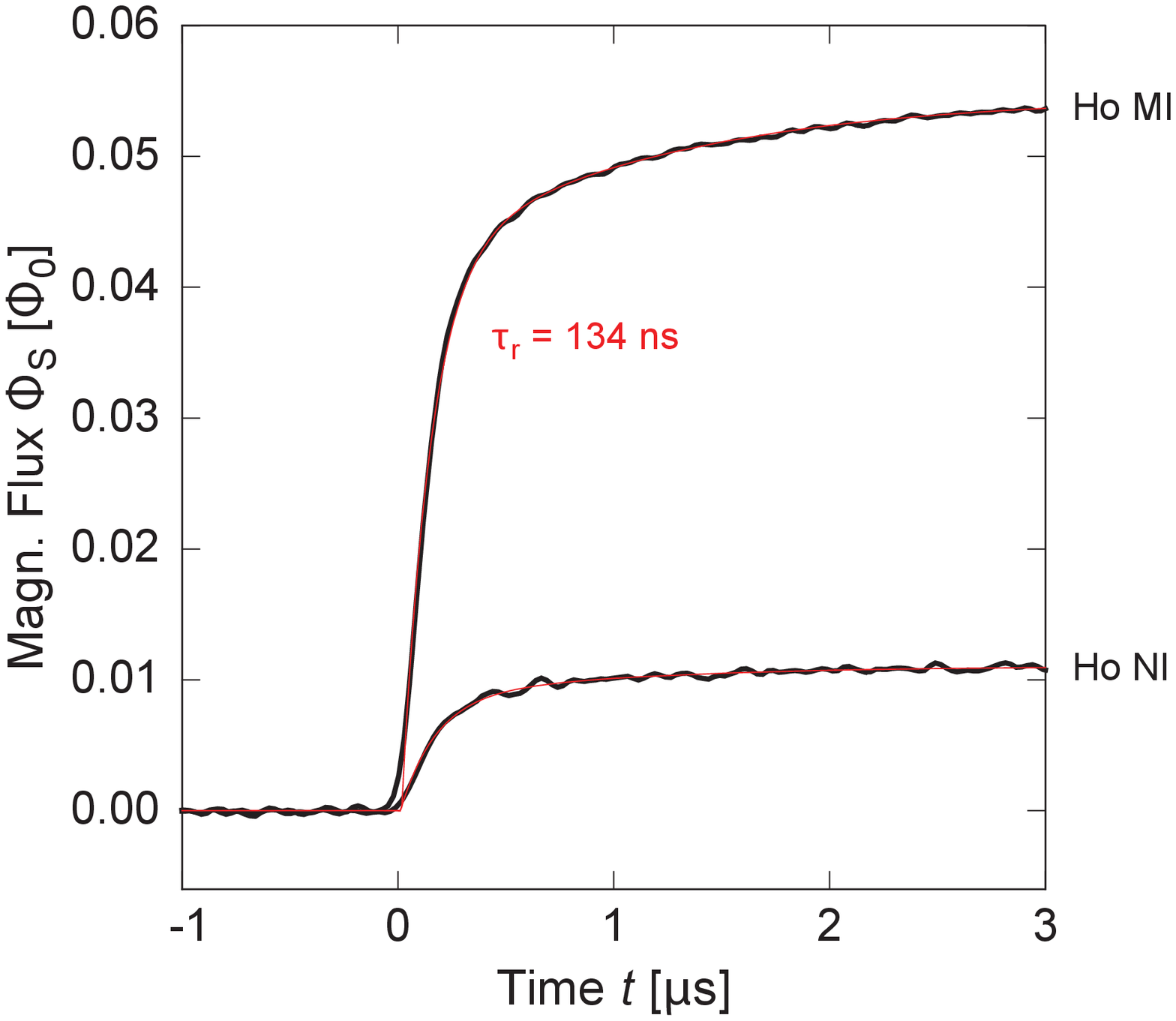}
%\caption{Method detail information}
%\label{fg:method_detail}
\end{center}
\endminipage\hfill
\caption{(Color online) a) Fit of the Mn $K_{\alpha}$ line. The energy resolution is $\Delta E_{\mathrm{FWHM}}\,\simeq\,7.6\,$eV. b) A rise time of $\tau_{\mathrm{r}}\,\simeq\,100\,$ns is extracted by the exponential fit.}
\label{performance}
\end{figure}

\noindent In a recent paper by Galeazzi et al. \cite{Galeazzi_2012}, several scenarios to reach the sub-eV sensitivity  have been described. A total statistics of $10^{14}-10^{16}$ counts in the full spectrum needs to be acquired, that leads to a required  $^{163}$Ho activity of more than $10^6$ Bq, and the detectors should meet the following requirements:

\begin{itemize}
\item[$\diamond$] fast signal rise-time, in order to reduce the un-resolved pile-up background due to the impossibility to de-convolve two or more events which happen in the same detector within a time interval shorter than the signal rise-time

\item[$\diamond$] energy resolution better than $\Delta E_{\mathrm{FWHM}}=10\,$eV to have the right accuracy in the end point description

\item[$\diamond$] possibility of multiplexing arrays of detectors without degrading the performance of the single pixel.
\item[$\diamond$] good linearity in order to precisely define the energy scale of the $^{163}$Ho EC spectrum and in particular fix the endpoint.
\end{itemize}
The ECHo experiment will be performed using
Metallic Magnetic Calorimeters (MMCs) \cite{Fleischmann_LTD13}. MMCs are energy dispersive
detectors typically operated at temperatures below $50\,$mK. These detectors consist of a particle absorber, where the energy is deposited, tightly connected to a temperature sensor which is weakly connected to a thermal bath. The deposition of energy in the absorber leads to an increase of the detector temperature. The temperature sensor of the MMCs is a paramagnetic alloy which resides in a small magnetic field. The change of temperature leads to a
change of magnetization of the sensor which is read-out as
a change of flux by a low-noise SQUID magnetometer. The
sensor material, presently used for MMCs, is a dilute alloy of erbium in gold,
Au:Er. The concentration of erbium ions in the sensor can be chosen to
optimize the detector performance and usually varies between 200
ppm and 800 ppm.
The spectral resolving power of a state of the art MMCs
for soft x-rays is above 3000. For completely micro-structured detectors, an energy resolution of $\Delta E_{\mathrm{FWHM}}\,=\,2\,$eV at $6\,$keV and a signal rise-time $\tau_{\mathrm{r}}\,=\,90\,$ns \cite{Pies_LTD14} have been achieved. Moreover the typical non-linearity at 6 keV is less than $1\%$ and the non-linear part can be described very well by a polynomial function of second order.
\noindent The achieved performance suggests that MMCs are suitable detectors for measuring the high precision and high statistics EC spectrum of $^{163}$Ho.
\noindent In order to perform a calorimetric measurement of the $^{163}$Ho EC spectrum, the $^{163}$Ho source has to be:
\begin{itemize}
\item[$\diamond$] part of the sensitive volume of detector to prevent the partial loss of energy
\item[$\diamond$] homogeneously distributed so that the detector response is position independent
\item[$\diamond$] completely contained in the detector in order to ensure a quantum efficiency for the emitted particles of $100\%$.
\end{itemize}
The optimum activity per pixel will be a compromise between having a high count rate per pixel and reduce the un-resolved pile-up as well as the heat capacity of the detector. Within our presently favored scenario the maximum activity per pixel can vary between 10 Bq and 100 Bq. The relatively low activity per pixel leads to a relatively large number of pixels to host the required activity of $10^6$ Bq. This requires the development of a multiplexed detector readout.
The read-out scheme for MMCs is compatible with several multiplexing techniques developed for low temperature micro-calorimeters, in particular with the microwave multiplexing which has the positive aspect to keep the performance of each detector in the array very close to that of a single pixel readout.

\begin{figure}[t]
\begin{center}
\includegraphics[width=0.7\linewidth, keepaspectratio]{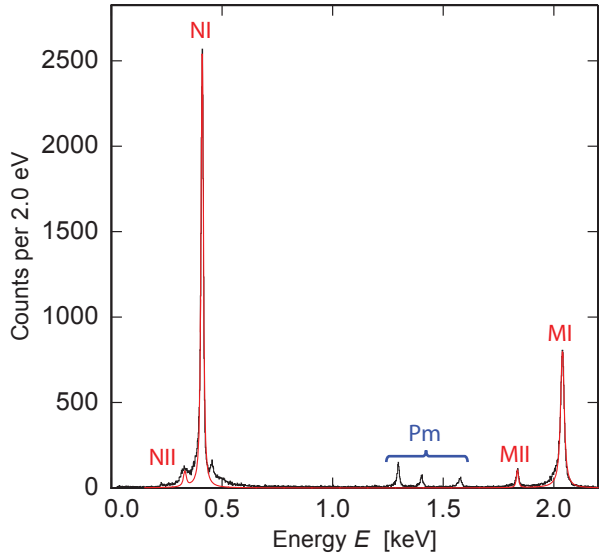}
\end{center}
\caption{(Color online) Calorimetrically measured $^{163}$Ho EC spectrum}
\label{spectrum}
\end{figure}

\noindent A first prototype detector chip consisting of four pixels having a gold absorber with implanted $^{163}$Ho has already been produced and tested \cite{Gastaldo_NIMA}. The ion implantation process was performed at ISOLDE-CERN \cite{kugler_2000}. The $^{163}$Ho ions have been implanted over a reduced area $160\times160\,\mu$m$^2$ of the first gold layer of the absorber having dimensions $190\,\times\,190\,\times\,5\,\mu$m$^3$. A second gold layer having as well dimensions of $190\,\times\,190\,\times\,5\,\mu$m$^3$ was deposited on top of the first layer. With this absorber fabrication, all the three mentioned requirements for embedding the source have been fulfilled \cite{Gastaldo_NIMA}. The $^{163}$Ho activity per pixel was about $10^{-2}$ Bq.
\noindent The first measurements  performed using only one pixel of the prototype chip showed that the implantation process did not degrade the performance of the MMC. An energy resolution of $\Delta E_{\mathrm{FWHM}}\,\simeq\,12\,$eV and the rise-time $\tau_{\mathrm{r}}\,\simeq\,100\,$ns have been measured and the non-linearity at 6 keV was less than $1\%$, as expected \cite{Ranitzsch_LTD14}. A second experiment had been performed with the same chip. Two pixels have been simultaneously measured for about two months. On one of them a $^{55}$Fe calibration source was collimated. Fig. \ref{performance} shows the performance achieved in this experiment.  An energy resolution of $\Delta E_{\mathrm{FWHM}}\,\simeq\,7.6\,$eV was achieved and the signal rise-time was $\tau_{\mathrm{r}}\,\simeq\,100\,$ns, as expected. Fig. \ref{spectrum} shows the spectrum obtained combining more than 30 files for each of the two pixels.
The results achieved using the MMC prototypes have shown that the requirements for the single pixel performance have been met. The ECHo collaboration aims at further improving the energy resolution of the MMC having $^{163}$Ho ions contained in the absorber and reach an energy resolution below $5\,$eV for multiplexed detectors.

%\begin{figure}
%\centering
%\begin{center}
%\includegraphics[width=0.40\linewidth]{Ka_combined.eps}
%\end{center}
%\hfill
%\includegraphics[width=0.40\linewidth, angle=-90]{rise_time_fit.eps}
%\caption{(Color online) a) Fit of the Mn $K_{\alpha}$ line. The energy resolution is $\Delta E_{\mathrm{FWHM}}\,\simeq\,7.6\,$eV. b) A rise time of $\tau_{\mathrm{r}}\,\simeq\,100\,$ns is extracted by the exponential fit.}
%\label{performance}
%\end{figure}

\section{MMC and microwave multiplexing}
\begin{figure}[t]
\minipage{0.45\textwidth}
\begin{center}
\includegraphics[width=1\textwidth]{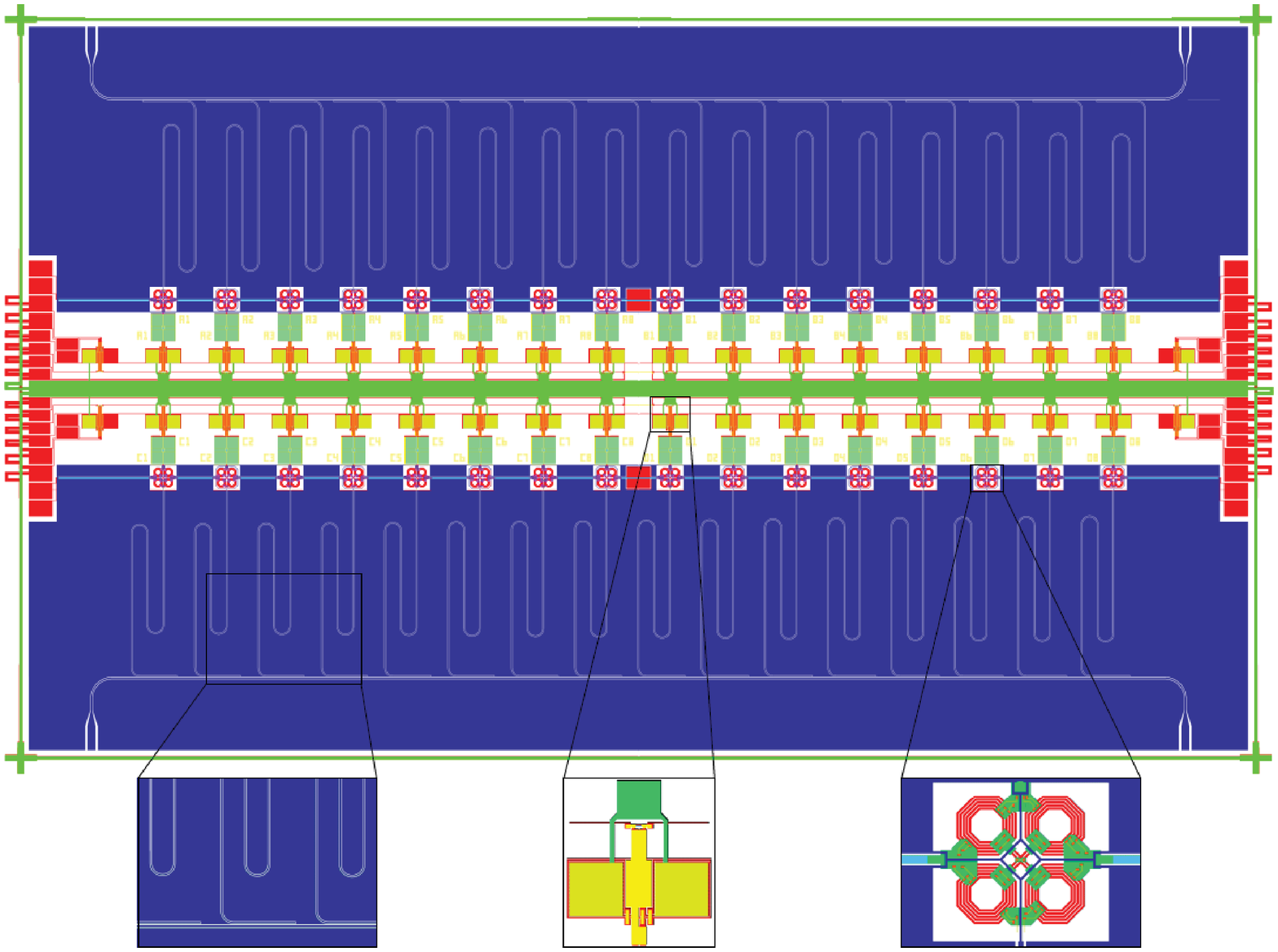}
%\caption{Methods}
%\label{fg:methods}
\end{center}
\endminipage\hfill
%\minipage{0.1\textwidth}
%\begin{center}
%\end{center}
%\endminipage\hfill
\minipage{0.5\textwidth}
\begin{center}
\includegraphics[width=1\textwidth]{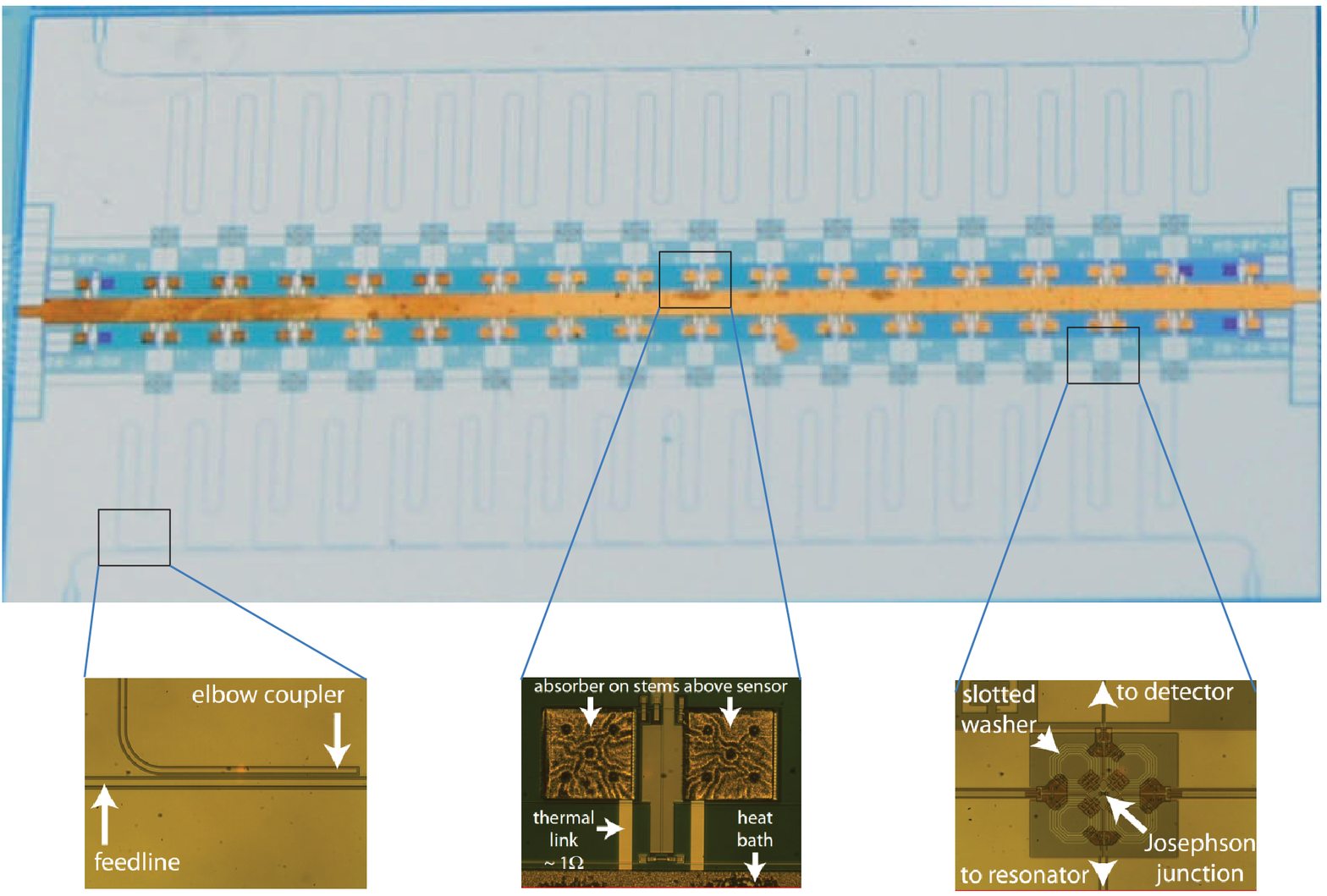}
%\caption{Method detail information}
%\label{fg:method_detail}
\end{center}
\endminipage\hfill
\caption{(Color online) a) Design of the 64-pixel MMC array with integrated microwave multiplexing readout. The magnifications show, from left to right, the elbow coupler of a resonator, the double meander design of the MMC and the rf-SQUID. b) One of the first micro-structured 64-pixel chip. The magnifications show,from left to right, the elbow coupler of a resonator, the double meander design of the MMC and the rf-SQUID.}
\label{mux_chip}
\end{figure}

Presently new chips consisting of 64 pixels which are read-out using the microwave multiplexing scheme \cite{Mates} have been developed. In this multiplexing scheme every detector is coupled to a non-hysteretic, un-shunted rf-SQUID which is coupled to a superconducting microwave resonator with high internal quality factor and unique resonance frequency. A change of magnetic flux inside the SQUID caused by an event in the detector leads to a change of the effective SQUID inductance and therefore, due to the mutual interaction, to a change of the resonance frequency of the corresponding microwave resonator. It is possible to measure the signal of each detector simultaneously by capacitively coupling the corresponding number of resonators to a common transmission line, injecting a microwave frequency comb driving each resonator at resonance and monitoring either amplitude or phase of each frequency component of the transmitted signal. Fig. \ref{mux_chip} shows the design of the 64-pixels chip and the micro-fabricated chip. A detailed description of single pixel geometry, resonator design and rf-SQUID properties as well as the expected performance of the detectors in the array is discussed in \cite{Kempf_LTD15}.

\section{$^{163}$Ho source: production and purification}
The production of a sufficient amount of $^{163}$Ho atoms in a radiochemically pure form is of paramount importance for the success of the ECHo experiment. Preliminary studies of different approaches to produce $^{163}$Ho have already been performed by the ECHo collaboration and by others \cite{Engle}. The production methods for $^{163}$Ho can mainly be divided into two branches:
\begin{itemize}
\item[$\diamond$] charged particle activation of suitable targets in reactions that lead to  $^{163}$Ho or its precursor $^{163}$Er which decays to  $^{163}$Ho (EC and $\beta ^+$ decay) with a half-life $T_{1/2} \,=\,75\,$min
\item[$\diamond$] thermal neutron activation of enriched $^{162}$Er targets.
\end{itemize}
The reaction that is typically used for the direct activation with proton beams is $^{\mathrm{nat}}$Dy$(p,xn)^{163}$Ho. Another possible direct reaction uses deuteron projectiles, $^{163}$Dy$(d,2n)^{163}$Ho.Examples for indirect $^{163}$Ho production are $^{\mathrm{nat}}$Dy$(\alpha,xn)^{163}$Er(EC, $\beta ^+$)$^{163}$Ho and $^{159}$Tb$(^7$Li$,3n)^{163}$Er(EC, $\beta ^+$)$^{163}$Ho.
 %For more details on the $^{163}$Ho production, please refer to the related sessions in this paper \ref{Susanta}\ref{Ulli}\ref{Engle}.
In the ECHo experiment all the described methods for production of $^{163}$Ho through charged particle activation processes as well as with neutron irradiation of a $^{162}$Er target are being considered, as well as possible new methods.
Typically, in the production of $^{163}$Ho, contaminants nuclides with decay properties that interfere with the desired clean $^{163}$Ho EC spectrum are produced. These include long-lived $\beta ^-$-decaying nuclides like $^{166 \mathrm{m}}$Ho. Thus, a careful separation of $^{163}$Ho is of paramount importance and will be performed by a combination of chemical and potentially physical separations before as well as after the irradiation of the sample. In this way, samples will be produced that contain contaminants on a level at which their contribution to the background of the calorimetric measurement is smaller then the intrinsic pile-up background.
 %A few processes have already been tested. A liquid-liquid extraction technique to separate erbium nuclides from the dysprosium target was performed on $\alpha$-irradiated natural dysprosium target at Saha Institute for Nuclear Physics in Kolkata. The developed chemical method is fast enough to complete the entire chemical process within one half-life of $^{163}$Er. On a $^{162}$Er enriched erbium sample which was irradiated for 11 days with thermal neutrons at the BER II reactor at the Helmholtzzentrum in Berlin (the thermal neutron flux is $\Phi = 1.3 \times 10^{14}\,$s$^{-1}$cm$^{-2}$) an ion-chromatography process using $\alpha$-hydroxyisobutyric acid was used to separate the holmium ions from the erbium target ions. This source after the purification step contains about $10^{16}$ $^{163}$Ho ions and will partially be used for the characterization of the contaminants as well as for the optimization of the separation and purification method. Moreover a large fraction of this source will be used for future detector tests as well for first experiments to determine the $Q_{\mathrm{EC}}$ value with Penning Traps.
Tests both at accelerator as well as at reactor facilities are already under way. The aim of these tests is to investigate and quantify the production of radioactive contaminants and to improve the purification methods in order to reach the required purity.
As a final step, the production of samples suitable for the calorimetric measurements will be optimized. A method that appears attractive include the formation of intermetallic samples \cite{Usolstev_2012} or ion-implantation.

\section{$^{163}$Ho EC spectrum parameterization}
In an Electron Capture process a nucleus $^A _Z$X decays by capturing an electron from the inner atomic shells and emitting an electron neutrino to $^A _{Z-1}$X. The daughter atom is left in an excited state. The atomic de-excitation is a complex process which includes cascades of both x-rays and electron emissions (Auger electrons and Coster-Kronig transitions). The possibility to measure all the energy released in the decay minus the energy taken away by the neutrino simplifies the description of the spectrum. The expected shape of the calorimetrically measured EC spectrum is:
\begin{equation}
\frac{dN}{dE_{\mathrm{C}}}\,=\,A(Q_{\mathrm{EC}}-E_{\mathrm{C}})^2\sqrt{1-\frac{m_{\nu}^2}{(Q_{\mathrm{EC}}-E_{\mathrm{C}})^2}}\sum{C_{\mathrm{H}}n_{\mathrm{H}}B_{\mathrm{H}}\,\phi^2_{\mathrm{H}}(0)\frac{\frac{\Gamma_{\mathrm{H}}}{2\pi}}{(E_{\mathrm{C}}-E_{\mathrm{H}})^2+\frac{\Gamma_{\mathrm{H}}^2}{4}}}	 \label{ECspectrum}
\end{equation}
and shows Breit-Wigner resonances centered at about the binding energy of the electron that was captured referred to the daughter atom nuclear potential, $E_{\mathrm{H}}$, where $H$ indicates the level from which the electron has been captured, with an intrinsic width $\Gamma_{\mathrm{H}}$. The intensities of these lines are given by the nuclear shape factors $C_{\mathrm{H}}$, the fraction of occupancy of the $H$-atomic shell $n_{\mathrm{H}}$, the squared wave-function of the captured electron calculated at the nucleus $\phi^2_{\mathrm{H}}(0)$ and a small correction, $B_{\mathrm{H}}$, due to electron exchange and overlap. The Breit-Wigner resonances are then modulated by the phase space factor which depends on the square of the electron neutrino mass $m(\nu_{\mathrm{e}})^2$ and the energy available to the decay $Q_{\mathrm{EC}}$. $A$ is a constant.
The purpose of the ECHo experiment is to improve the theoretical description of the EC spectrum of $^{163}$Ho, in particular by investigating the modification due to the environment. This will be done by applying the complete screening
approximation of Density Functional Theory (DFT) \cite{HOHE64},\cite{EGEL87} to calculate the energy level of electrons and by improving the calculation of the partial decay rates by considering
the finite size of the nucleus. This can be done by calculating the corresponding nuclear matrix elements
for the transition $^{163}\mathrm{Ho}(7/2^-) \rightarrow ^{163}\mathrm{Dy}(5/2^-)$ and
capture of $s_{1/2}$ and $p_{1/2}$ electrons using the Quasi-Particle Random Phase Approximation (QRPA) approach \cite{QRPA}.
The total energy available for the decay of $^{163}$Ho to $^{163}$Dy is a fundamental parameter to reach the sub-eV sensitivity to the electron neutrino mass by the analysis of the high energy part of the $^{163}$Ho EC spectrum. The recommended value is $Q_{\mathrm{EC}}\,=\,2.555\,(16)\,$keV as can be found in \cite{QEC}, but other measurements give values that range from about 2.3 keV \cite{Andersen_Phys_lett_1982} to about 2.8 keV obtained by two calorimetric measurements performed using low temperature detectors \cite{Ranitzsch_LTD14} and \cite{Gatti_1997}. The aim of the ECHo collaboration is to reach a precision on the $Q_{\mathrm{EC}}$ of 1 eV or better. After preliminary measurements at the double-Penning trap mass spectrometers SHIPTRAP \cite{SHIPTRAP} and TRIGA-TRAP \cite{TRIGA}, the final high accuracy measurement will be accomplished by the novel Penning-trap mass spectrometer PENTATRAP \cite{Repp}, \cite{Roux}. The uniqueness and complexity of this Penning-trap mass spectrometer is associated with the unprecedented relative accuracy of a few parts in $10^{12}$ with which the $Q_{\mathrm{EC}}$ must be measured.

\section{Conclusion}
The ECHo experiment has the aim to investigate the electron neutrino mass in the sub-eV range. The performance achieved by the first prototype of MMCs with $^{163}$Ho implanted in the gold absorber demonstrated that MMCs fulfill all requirements on the single pixel level. We presented the different aspects that are presently under investigation within the ECHo project. The next goal of the ECHo collaboration is to set a small scale experiment based on arrays of MMCs with embedded $^{163}$Ho ions, for a total pixel number of about 100, read out by the microwave multiplexing scheme. Within this experiment about $10^{10}$ events will be collected. The analysis of the end part of the spectrum combined with the precise knowledge of the parameters describing the spectrum will allow to reach the sensitivity of few eV on the electron neutrino mass. A full scale experiment holds the prospect of reaching the sub-eV sensitivity.

%\begin{acknowledgements}

%\end{acknowledgements}

%\pagebreak

\end{document}